\documentclass[pra,twocolumn,floatfix]{revtex4}

\mathchardef\ogon="012C%
\newcommand{\as}{a\kern-0.22em\lower.40ex\hbox{$_{\ogon}$}}
\newcommand{\As}{A\kern-0.22em\lower.40ex\hbox{$_{\ogon}$}}
\newcommand{\es}{e\kern-0.22em\lower.40ex\hbox{$_{\ogon}$}}
\newcommand{\Es}{E\kern-0.22em\lower.40ex\hbox{$_{\ogon}$}}

\newcommand{\NN}{{\cal{N}}}
\newcommand{\KK}{{\cal{K}}}

\newcommand{\RR}{{\cal{R}}}
\newcommand{\QQ}{{\cal{Q}}}

\newcommand{\xx}{{\bf x}}

\newcommand{\pp}{{\bf p}}

\usepackage{epsfig}
\usepackage{color}
\usepackage{amssymb,amsmath}

\begin{document}

\title{Correlation functions of cold bosons in an optical lattice}
\author{Radka Bach}
\author{Kazimierz Rz{\as}{\.z}ewski}
 \altaffiliation[Also at ]{Cardinal Wyszy{\'n}ski University, al.\ Lotnik{\'o}w 32/46, 02-668 Warsaw, Poland}
\affiliation{Center for Theoretical Physics, Polish Academy of Sciences, al.\ Lotnik{\'o}w 32/46, 02-668 Warsaw, Poland}
\date{\today}

\begin{abstract}
We investigate the experiment of collapses and revivals of matter wave field in more detail.
To this end we calculate the lowest-order correlation functions of the Bose field. We compare predictions of the
total Fock state with the commonly used coherent state approximation. We also show how to observe an interference pattern
for the celebrated Mott state.
\end{abstract}

\maketitle

\section{Introduction}
\label{sec:intro}

Cold bosons in optical lattices have gained a lot of attention recently because of the ease and precision 
with which they can be manipulated \cite{Immanuel:PhysicsToday}. For example, it was an optical lattice that enabled
the observation of collapses and revivals of a matter wave field \cite{Immanuel}. 
Even though the theoretical analysis of this experiment provided by
authors is based on the notion of coherent states and as such is---in principle, at least---inadequate to describe
systems with total number of atoms fixed, it seems to work quite well. It is our purpose to describe this experiment more
accurately, in particular to examine the role of
total number of atoms conservation and identify situations in which it will clearly manifest itself.

This will be achieved with the aid of correlation functions. Contrary to the common belief, there is no need to build
a separate experimental set-up to measure higher-order correlation functions in atomic systems -- it is enough to 
analyze photographs of a cloud of atoms. Such photographs, which are typically obtained in the final stage of
experiments with cold gases, are nothing else but a simultaneous detection of many atoms and therefore probe
the correlation function of the order of the number of atoms \cite{Javanainen}. And since the observed system
comprises of a fixed number of atoms, it is described by a Fock state; consequently it is possible to reconstruct
low-order correlation functions out of these measurements \cite{Bach}. Having the possibility of measuring correlation
functions, it is justified to investigate them theoretically as well.

The article is organized as follows. In Section \ref{sec:experiment} we start with a brief description of the experiment in
which collapses and revivals of a matter wave field were observed. On this basis two states of the system are
introduced: one that obeys the total number of atom conservation and one that violates it (the latter for
		comparison). Correlation functions corresponding to both states are calculated and investigated
in Section \ref{sec:corrfun}. Inter alia, we identify situations in which predictions of the two examined states
differ. We also note that in the collapsed phase the system effectively behaves as if there was no
site-to-site coherence, as pointed out in Section \ref{sec:random}. Even then, though, 
the second-order correlation function shows nontrivial structure and interference pattern should be seen in a 
single photograph. Counter-intuitively, this is also the case for Mott insulator, as shown in Section 
\ref{sec:mott}, and it is extremely difficult to tell these states apart on the basis of a single or many measurements.
We conclude in Section \ref{sec:conclusions}.

\section{The experiment}
\label{sec:experiment}

The experiment in which collapses and revivals of the matter wave field were observed \cite{Immanuel} was
composed of a few steps. After preparing a Bose-Einstein condensate in a harmonic oscillator potential, an optical
lattice potential was slowly raised. The height of this lattice has been chosen such that the system was
still completely in the superfluid regime. Then, the intensity of light creating the lattice was rapidly increased
and the resulting optical lattice was so high that the tunneling between the sites was strongly suppressed. Was this
raising done adiabatically, the system would move to the Mott insulating phase; but because it was done rapidly,
the atom number distribution at each well from a superfluid state was preserved at the high lattice potential,
	thus producing a mixture of Fock states with different number of atoms at each well.
The system was then left to evolve for some time, which was varied from experiment to experiment 
(we will call it interaction time and denote $T$). Finally, after switching off all potentials the atomic cloud
was allowed to expand freely for time $t$ before shooting a photograph. As a function of the interaction time $T$
a collapse and a revival of the interference pattern were seen.

We are going to describe this experiment via a simple model which nonetheless includes the most relevant features
of the system. For example, we will neglect the non-uniformity of the optical lattice stemming from additional
harmonic oscillator potentials (one used to create the Bose-Einstein condensate and another one
associated with widths of laser beams), but consider a system composed of $\NN$ atoms distributed among $\KK$ equivalent 
lattice sites instead. The notion of atom density, $\rho =\NN/\KK$, will be used interchangeably. 
The wells' exact shape is also not important since we are going to assume that atoms are so cold that they do not
have enough energy to occupy states other than the ground state, which is justified by the fact
that the most important part of the evolution---the one from which collapses and revivals stem---
takes place when lattice potential is very high and when the energy of an atom is far less than the energy gap to the
first excited state.

Prior to turning on the strong optical lattice the system was in a superfluid state and so right after raising the
potential the distribution of atoms between wells is multinomial. Hence the state of the system is:
\begin{equation}
\begin{split}
\left| \psi (T=0) \right\rangle &= 
\sqrt{ \frac{\NN!}{\KK^\NN} }
\underbrace{ \sum_{n_1=0}^\NN \sum_{n_2=0}^\NN \cdots \sum_{n_\KK=0}^\NN }
\\ &
\frac{1}{\sqrt{n_1! n_2! \ldots n_\KK!}}  \: 
\left| n_1, n_2, \ldots, n_\KK \right\rangle 
\end{split}
\end{equation}
where the underbrace denotes total number of atoms conservation law, i.e. 
$n_1 + n_2 + \cdots + n_\KK = \NN$ and we have already assumed that each atom can occupy any of the wells with equal
probability.

Once prepared, the system is left to evolve for time $T$. Since atoms are now imprisoned in a strong
optical lattice in which tunneling is highly suppressed, in each well they evolve independently of others. Assuming
contact interactions between atoms, the Hamiltonian in each well is effectively of the form:
\begin{equation}
\hat{H} = \frac{\hbar g}{2} \hat{n} \left( \hat{n} -1 \right)
\end{equation}
where $\hat{n}$ is the operator of the number of atoms in this well and $g$ denotes the rescaled coupling constant:
\begin{equation}
g = \frac{4 \pi \hbar a}{m} \int \! {\mbox{d}} \xx \left| \varphi(\xx) \right|^4
\end{equation}
($\varphi(\xx)$ is the wavefunction of the ground state of a well and $a$ is the $s$-wave scattering length). 
Then, after time $T$ the state of the system is:
\begin{equation}
\begin{split}
	\left| \psi_{\mbox{\footnotesize mnm}}(T) \right\rangle &=
	\sqrt{\frac{\NN!}{\KK^\NN}} \underbrace{ \sum_{n_1=0}^\NN \sum_{n_2=0}^\NN\cdots \sum_{n_\KK=0}^\NN } 
	\\ & 
   c_{n_1}(T) \cdots c_{n_\KK}(T) 
	\left| n_1, n_2, \ldots, n_\KK \right\rangle
\end{split}
\end{equation}
where the coefficients $c_{n}(T)$ are:
\begin{equation}
c_{n}(T) = \frac{1}{\sqrt{n!}} \exp\left\{ - {\mbox{i}} \frac{gT}{2} n(n-1) \right\}.
\end{equation}

To investigate the role of total number of atoms conservation law, however, we are going to analyze a coherent
counterpart of this state as well:
\begin{equation}
\begin{split}
\left| \psi_{\mbox{\scriptsize coh}}(T) \right\rangle &=
{\mbox{e}}^{-\rho\KK/2}  \sum_{n_1=0}^{\infty}
\sum_{n_2=0}^{\infty} \cdots \sum_{n_\KK=0}^{\infty} 
\alpha^{n_1+n_2+\cdots+n_\KK} 
\\ &
c_{n_1}(T) \cdots c_{n_\KK}(T) 
\left| n_1, n_2, \ldots, n_\KK \right\rangle 
\end{split}
\end{equation}
where the coefficients $c_n(T)$ are defined as before and $|\alpha|^2 = \rho$ for consistency. 

As far as a subsystem composed of a fixed number of wells is concerned, the multinomial state approaches the coherent
one when the number of atoms and the number of wells are increased in such a way that the density is kept constant 
\footnote{At least from the point of view of the correlation functions we are investigating.}. In this
limit---let us call it thermodynamic limit in analogy with statistical mechanics---the remaining part of the system
serves as a particle reservoir and not surprisingly the distribution of atoms factorizes and becomes
poissonian in each well independently, as it is for coherent states. 

\section{Correlation functions}
\label{sec:corrfun}

Let us now calculate explicit analytical formulas for correlation functions for the multinomial
and coherent state introduced above. The $r$-th order correlation function is defined as:
\begin{equation}
\begin{split}
& G^{(r)} (\xx_1,\xx_2,\ldots,\xx_r;T,t) = 
\\ 
& \left\langle \psi(T) \left| \hat{\Psi}^\dagger(\xx_1,t) \cdots \hat{\Psi}^\dagger(\xx_r,t) 
\hat{\Psi}(\xx_r,t) \cdots \hat{\Psi}(\xx_1,t)
\right| \psi(T) \right\rangle
\end{split}
\end{equation}
In the above formula two distinct times appeared: the interaction time $T$, denoting how long atoms were left to
evolve in the strong optical lattice potential, and the measurement time, $t$, which is
the time between switching off all binding potentials and shooting the actual
photograph. During the latter time there are no external potentials and we also assume that the interaction does
not play any major role---consequently the only effect this process has on the system is the free expansion of wells'
wavefunctions. Although this way the interaction phases that atoms might have acquired are neglected,
the assumption is not only justified (since during expansion the system becomes extremely dilute), but it also 
clarifies the overall picture. Each of the times is now responsible
for different physical effects: the interaction time, $T$, governs the phases of different Fock states and therefore
is responsible for collapses and revivals, while the measurement time, $t$, determines the
behaviour of wells' wavefunctions and as such does not influence the relation between different Fock states. 
Note also that the time $t$ is treated here rather as a convenient control parameter than a true argument of the
correlation function.

The field operator $\hat{\Psi}(\xx,t)$ can be decomposed in any complete basis of annihilation operators.
In the system we are considering it is convenient to introduce annihilation operators of atoms
at individual sites' ground states:
$ \hat{\Psi}(\xx,t) = \sum_{k=1}^\KK \varphi_k (\xx,t) \hat{a}_k$. Then after some lengthy though straightforward
calculations one obtains:
\begin{equation}
\begin{split}
& G^{(r)}(\xx_1,\ldots,\xx_r; T,t) = \rho^r \sum_{l_1,\ldots,l_r} \sum_{k_1,\ldots,k_r}
\omega(\{l,k\},T)
\times \\  
& \times \varphi^*_{l_1}(\xx_1,t) \cdots \varphi^*_{l_r}(\xx_r,t) 
\varphi_{k_r} (\xx_r,t) \cdots \varphi_{k_1} (\xx_1,t) \;
\label{eq:finalformulas}
\end{split}
\end{equation}
where:
\begin{equation}
\omega_{\mbox{\footnotesize coh}}(\{l,k\},T) = 
\RR(\{l,k\},T) \exp\left\{ \rho \QQ(\{l,k\},T) \right\}
\end{equation}
for coherent states and 
\begin{equation}
\begin{split}
& \omega_{\mbox{\footnotesize mnm}}(\{l,k\},T) = \\
& \frac{\NN!}{\NN^r (\NN-r)!} \RR(\{l,k\},T)
	 \left( 1+ \frac{\QQ(\{l,k\},T)}{\KK}\right)^{\NN-r}
\end{split}
\end{equation}
for multinomial states. The functions $\RR$ and $\QQ$ are:
\begin{subequations}
\label{eq:finalformulasRQ}
\begin{align}
& \RR(\{l,k\},T)    =
\\ & =
 \exp\left\{{\mbox{\footnotesize i}} \frac{g T}{2} \sum_{i=1}^{\KK} \left[ 
	 \left( \sum_{j=1}^r \delta_{l_j,i} \right)^2 - \left( \sum_{j=1}^r \delta_{k_j,i} \right)^2 
\right] \right\}
\nonumber \\ 
&	\QQ(\{l,k\},T) =
\sum_{i=1}^{\KK} \left( {\mbox{e}}^{{\mbox{\footnotesize i}} g T 
		\sum_{j=1}^r \left( \delta_{l_j,i} - \delta_{k_j,i} \right) } -1 \right)
\end{align}
\end{subequations}
First of all the expression for the correlation does not depend on the interaction time $T$ nor the coupling strength
$g$ alone, but on the product of the two. It means that at $T=0$ the interacting system effectively mimics the behaviour of an
ideal gas, for which, as can be easily checked, all correlation functions factorize (for multinomial states there
appears an additional proportionality constant). %$\NN!/ \NN^r (\NN-r)!$).
And because the functions $\RR$ and $\QQ$ are
periodic functions of $gT$, this also happens for any time such that $gT=2 k\pi$, where $k$ is an integer. 
This re-appearance of the interference pattern is nothing else but the revival phenomenon.

\subsection{The first-order correlation function}

Density---the first-order correlation function---is obtained from above formulas via setting $r=1$ and can be written
as:
\begin{equation}
\label{eq:density}
G^{(1)} (\xx) = 
\rho \sum_{k} \left| \varphi_k(\xx) \right|^2 +
\kappa \rho
\sum_{l} \sum_{k \neq l} \varphi_l^*(\xx) \varphi_k(\xx)
\end{equation}
where:
\begin{equation}
\kappa = \left\{
	\renewcommand{\arraystretch}{1.2}
	\begin{array}{lcl}
   \displaystyle
\exp \left\{ - 4 \rho \sin^2 \left(\frac{g T}{2} \right) \right\}
	& & {\mbox{for coherent states}} \\
	\displaystyle 
\left[ 1-\frac{4}\KK \sin^2\left(\frac{g T}{2} \right) \right]^{\NN-1}
	& & {\mbox{for multinomial states}} 
	\end{array} 
	\renewcommand{\arraystretch}{1}
	\right.
\end{equation}
First of all, the density is a sum of two kinds of terms: the
background, which is a simple consequence of having on average $\rho$ atoms at
each lattice site, and interference terms. The presence and relative amplitude of the latter is governed by the 
coefficient $\kappa$. 

\begin{figure}[htb!]
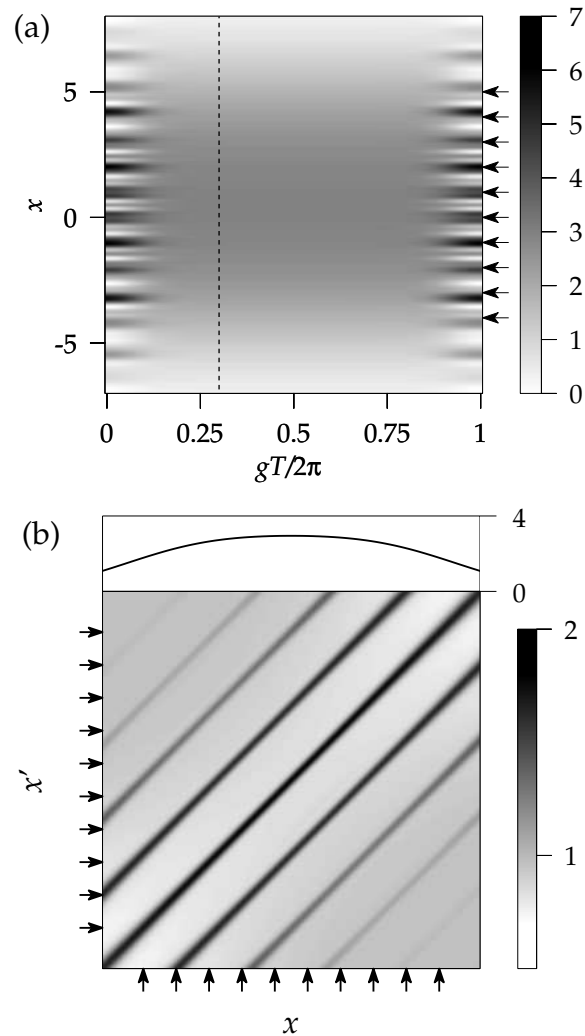

\begin{center}
\epsfig{file=art_fig_g1.ps} \\
\epsfig{file=art_fig_g2.ps}
\end{center}
\caption[fig:typicalg2]{First- and second-order correlation functions for one-dimensional system 
	composed of $\KK=10$ wells and $\NN=30$ atoms. The wells' wavefunctions were Gaussians with 
	initial width, $\sigma=0.12$ (this particular value has been chosen such that the resulting 
	nonorthogonality of neighbouring wells' 
	wavefunctions is still negligible---of order of $10^{-31}$---but there are only few fringes in the second-order
	correlation function) and the measurement time was $t=3\sigma$. Arrows denote the positions of wells' centers.
	{\bf (a)} Density profiles versus the interaction time, $gT$. Both the collapse and the revival of interference pattern
	is clearly visible.  
	{\bf (b)}: Close-up for a certain value of interaction time, $gT = 0.3 \times 2 \pi$.
	The curve at the top is the density profiles (this is a cut along the dotted line from the figure above). 
   The 2-dimensional density plot denotes the normalized second-order correlation function, 
	$g^{(2)}(x,x')$, where $x$ and $x'$ are plotted at the vertical and horizontal axis, respectively. }
\label{fig:typicalg2}
\end{figure}

The Fig.\ \ref{fig:typicalg2}(a) depicts typical density profiles versus the interaction time.  To observe
interference pattern two conditions must be simultaneously fulfilled: (i) the interaction time $gT$ is such that the
coefficient $\kappa$ is not 
vanishing and (ii) the measurement time, $t$, is such that the wavefunctions $\varphi_i(\xx)$ substantially overlap.
The latter condition controls the exact shape of interference pattern observed while the first one influences its
visibility.  
To compare predictions of multinomial versus coherent states it is therefore enough to investigate the coefficient
$\kappa$ only.  It is astonishing how quickly---when increasing the number of wells---predictions of the two states 
coincide: practically from 3 wells upwards there is no way to distinguish multinomial from coherent states through 
density profiles.
Only for two wells, $\KK=2$, is there a substantial discrepancy: binomial distribution predicts 
an additional revival at $gT=\pi$.
Interesting enough, the interference pattern observed at this additional revival is in-phase with the pattern
at $gT=0$ for odd number of atoms and out-of-phase for even number of atoms. One might think that
investigating the situation with only two wells 
is pushing the analysis beyond the limits set by experimental reality. This is not true,
however. Only slight modification of the experiment is required to go into this interesting regime \footnote{This idea
	originates from private communication with Immanuel Bloch}. 
Imagine that before the first optical lattice (the one, in which the atomic system is in a
superfluid state) is turned on, another optical
lattice is raised: a lattice deep enough to force the system into the Mott insulator state, with a
well defined number of atoms per site. Let us now turn on a lattice with half the spatial period of the
one already existing: this will produce a binomial distribution of atoms in each deep potential well
between two shallow wells. 

The vanishing of the coefficient $\kappa$ for almost all interaction times apart from the ones 
close to revival times is the origin of the collapse phenomenon.
To estimate the time of collapse $T^*$ one can 
require the coefficient $\kappa$ to reach a pre-set small value, $\epsilon$. Then:
\begin{subequations}
\begin{align}
\cos \left( g T^*_{\mbox{\scriptsize mnm}} \right) & \leq 
1-\frac{\KK}{2} \left( 1 - \epsilon^{\frac{1}{\NN -1}} \right) 
\\
\cos\left( g T^*_{\mbox{\scriptsize coh}} \right) & \leq  
1 + \frac{\ln \epsilon}{2 \rho}
\end{align}
\end{subequations}
Note that the collapse is (apart from $\KK=2,3$ cases) never exact, i.e.\ when demanding the interference fringes to
vanish completely, which corresponds to setting $\epsilon=0$, the above inequalities cannot be satisfied. At least
not without simultaneously increasing the density $\rho$. However, for any practical purposes it is enough to have 
the visibility of fringes smaller than the sensitivity of detectors to speak about complete collapse of interference
pattern. For small number of wells, on the other hand, there exist interaction times for which fringes vanish
identically: these are $g T^* = \frac{1}{4} \times 2 \pi$ and $g T^* = \frac{3}{4} \times 2 \pi$ for $\KK=2$ case,
and $g T^* = \frac{1}{3} \times 2 \pi$ and $g T^* = \frac{2}{3} \times 2 \pi$ for $\KK=3$ case.

\subsection{The second-order correlation function}
Let us concentrate on the normalized second order correlation function, defined as:
\begin{equation}
g^{(2)}(\xx,\xx') = \frac{G^{(2)}(\xx,\xx')}{G^{(1)}(\xx) G^{(1)}(\xx')}
\end{equation}
because this way information about the density is set aside and only correlations are being investigated. 
For interaction time such that $gT=2 k \pi$, where $k=0,1,2,\ldots$, i.e.\ exactly the revival time,
the normalized second-order correlation function is constant, because then the ideal gas case
is recovered and consequently:
\begin{equation}
g^{(2)}_{gT=2k\pi}(\xx,\xx') = \left\{ \begin{array}{lcl}
	1 & & \mbox{for coherent state} \\
   1-\frac{1}{\NN} & & \mbox{for multinomial state} 
	\end{array} \right.
\label{eq:g2_background}
\end{equation}
Already, there is a difference between the two states examined: multinomial states predict a small antibunching effect,
$g^{(2)}(\xx,\xx')<1$, which is due to the total number of atoms conservation law: 
the chance of detecting an atom at position $\xx$ is smaller if another atom was already detected somewhere else 
(than if it was not) because there are less atoms altogether. Note that this result will hold
also for other interaction times, provided $\xx$ and $\xx'$ are distant enough.

However, if one moves away from the exact revival time, so that $gT$ is no longer an integer multiple of $2 \pi$, an
additional structure in the second order correlation function appears. Typically, these are diagonal stripes in the
$x-x'$ plane, such that the normalized second order correlation function depends only on the difference of its
spatial arguments: $g^{(2)}(x,x') \approx g^{(2)}(x-x')$, as seen at Fig.\ \ref{fig:typicalg2}(b) (at least far from the
boundaries of the system). It means that in every realization interference fringes are going to be observed, 
but the whole pattern will move from shot to shot and will vanish after averaging,
producing a smooth density profile. 
The range of $g^{(2)}$ is closely related to the size of expanding wavepackets. Speaking roughly, 
to have a nontrivial correlation,
wavepackets must spread over the distance equal to the separation between the points at which $g^{(2)}$ is calculated
	or measured.
The separation between the fringes, on the other hand, is determined by the initial size of expanding wavepackets, 
or, in other words, by the
momenta involved: the steeper the optical lattice potential, the more localized the wavefunction and the more fringes
in the second-order correlation function structure.

\begin{figure}
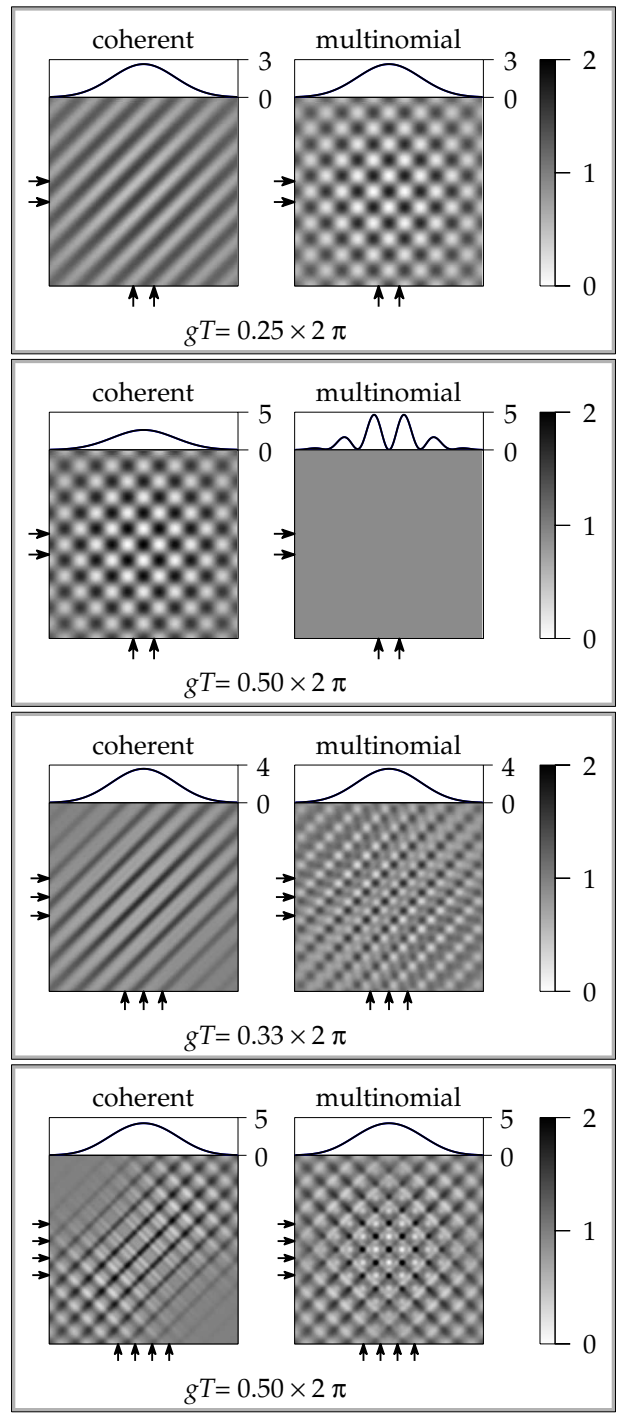

\centering
\epsfig{file=art_fig_magic_a.ps} \\
\epsfig{file=art_fig_magic_b.ps} \\
\epsfig{file=art_fig_magic_c.ps} \\
\epsfig{file=art_fig_magic_d.ps} \\
\caption[Magic times]{Situations, at which coherent and multinomial states give distinctively different predictions.
Density plots denote the normalized second-order correlation function, $g^{(2)}(x,x')$, where $x$ and $x'$
are on the horizontal and vertical axis, respectively. Arrows denote the position of the wells' centers. 
One-dimensional graph on top of the density plot is the density, $G^{(1)}(x)$. 
The parameters of physical systems, apart from the number of wells and the interaction time, are identical for all
plots: atom density, $\rho=5$, measurement time, $t=2\sigma$, and initial width of expanding wavepackets, $\sigma=0.12$. }
\label{fig:magic}
\end{figure}

The formulas for correlation functions are quite distinct for both kinds of states, it is thus
somewhat surprising that the final plots are practically indistinguishable. 
To spot possible discrepancies we have investigated
coefficients $\omega(\{l,k\})$ defined by Eq.\ (\ref{eq:finalformulas}) in more detail,
and found that even though these coefficients depend on four indices, they can take only one of six values
(this happens---technically speaking---because functions $\RR$ and $\QQ$ depend not on
the values of the indices themselves, but on their mutual relation, i.e.\ on the way they split into equivalence
classes with respect to the number of identical values). In the limit $\KK\rightarrow \infty$ and
$\NN\rightarrow \infty$ such that $\rho=\NN/\KK={\mbox{const}}$, the coefficients corresponding to multinomial states
become identical with the ones stemming from coherent states, which is a manifestation of the equivalence of
the two states in the thermodynamic limit. This limit is achieved so quickly that
from $\KK=5$ upwards there is practically no way of distinguishing between
the two states \footnote{Provided that the density is not extremely low, i.e.\ the number of atoms, $\NN$, is at least 
of order $10$ rather than of order $1$.}. 
Only for $\KK=2,3,4$ wells case is there a room for discrepancies and indeed such situations are found and depicted in
Fig. \ref{fig:magic}.

The conclusions presented above---stemming from one-dimensional simulations---are not altered
when extending the analysis to higher dimensions. With one important note, however.
The structure in the second-order correlation function implies that interference pattern should be seen in 
every single realization and smooth out when averaged over many experiments, producing a relevant density profile. 
This is true for a measurement
which detects atoms in the full, 3-dimensional physical space and is in general false for its 2-dimensional projection
(think about holograms versus photographs). The latter is governed by a column-averaged
correlation function, in which information about correlations in the direction of the illuminating light pulse is
lost, and such a situation corresponds to performing many 2-dimensional experiments and averaging over them. 
This explains why the collapse was seen at all in a single experimental run. Numerical simulations of a lattice
composed of $5\times5$ sites show that the visibility of the interference pattern drops from 44\% when averaged 
over one layer to 19\% when averaged over all 5 layers, and the effect is expected to be even more profound for
larger systems.

\section{Incoherent systems}
\label{sec:random}

The system examined herein before was implicitly assumed to be perfectly coherent, in a sense that in numerical
simulations wavefunctions of all optical lattice sites had identical phases. This was because in the experimental
situation the system prior to the turning on the strong optical lattice potential was in a superfluid phase, which
exhibits long-range coherence. However, in principle, the same formalism is also applicable to the situation in 
which there is no long-range coherence in the initial state, i.e.\ the wavefunctions' phases are completely random.
Even though the physical relevance of such a state might be questionable, it is worth to examine this limit as well,
because then one could separate effects originating from having a mixture of Fock states at each site from the ones
necessarily requiring the coherence between sites. The state with completely random phases is also interesting
because---at least when large measurement times are concerned---it can be to a large extent investigated 
analytically independently of its size and dimension.

Let us therefore assume that (i) the wavefunctions of individual wells have random phases: $\varphi_k(\xx) \rightarrow
{\mbox{e}}^{{\mbox{\footnotesize i}} \phi_k} \varphi_k(\xx)$, where $\phi_k$ are random numbers between $0$ and $2 \pi$,
and (ii) the measurement time $t$ is so large that at each point in space wavefunctions of many sites overlap. Then, 
in the expressions for the first- and second-order correlation
functions all terms that explicitly depend on phases $\phi_k$ will average out, producing:
\begin{subequations}
\label{eq:random}
\begin{align}
\overline{G^{(1)}} (\xx) &= \rho \sum_{k} \left| \varphi_k(\xx) \right|^2 
\\
\overline{G^{(2)}} (\xx,\xx') &=
\omega_0 \overline{G^{(1)}}(\xx) \overline{G^{(1)}}(\xx')
+ \nonumber \\ & 
+ \omega_0 \rho^2 \sum_{l} \sum_{k \neq l} 
\varphi_{l}^*(\xx) \varphi_k(\xx) \varphi_{l}(\xx') \varphi_{k}^* (\xx') 
\end{align}
\end{subequations}
where $\omega_0=1$ for coherent and $1-\frac{1}{\NN}$ for multinomial states.

Note that these expressions do not depend upon the interaction time at all! 
Therefore the fact that the experimentally observed pattern changed with interaction time is a clear sign
of the coherence in the initial state. The density shows no interference pattern and is exactly the same as the one of 
system with coherence in the collapsed phase, i.e.\ when $\kappa$ in Eq.\ (\ref{eq:density}) is close to zero.
Note however that here the mechanism is intrinsically different: 
The interference pattern is absent not because of the collapse of the matter wave field at each individual
well, but because of the washing-out effect when overlapping subsystems with different phases. If one looked
at a single well subsystem, one would have seen collapses and revivals there, just as it was pointed out in the
first article concerning this phenomenon in Bose-Einstein condensates \cite{Wright}. 

The second-order correlation function, on the other hand, shows additional structure, which
depends on the overlap of wells wavefunctions, $\varphi_{l}^*(\xx) \varphi_k(\xx)$. If the measurement time
is large enough this function is just a smooth profile modulated by a spatially-dependent phase of
the form $\exp( {\mbox{i}} \pp \xx )$, where $\pp$ is related to initial widths of expanding wavepackets, and then
the pattern observed in the second-order correlation function comprises of diagonal stripes in the
$\xx-\xx'$ hyperplane, just as the ones seen at Fig.\ \ref{fig:typicalg2}(b). 
The fact that in the collapsed phase of the evolution systems with perfect coherence have practically the same
first- and second-order correlation functions as systems with completely random phases implies that it is impossible
to distinguish between them, neither in a single nor many experiments. In that sense one could say that in the
collapsed phase the system with long-range coherence mimics the behaviour of the system with no coherence at all.

The difference between multinomial and coherent states manifests itself only in the coefficient
$\omega_0$ ($1$ versus $1-\frac{1}{\NN}$), which leads to a small antibunching effect, as also seen in the
case of systems with perfect coherence. Similarly, the structure in the second-order correlation function vanishes 
when column averaged. This is a
consequence of the fact that the free evolution propagator acts independently in each direction.
Initially the wavefunctions were very well localized at different points in space and therefore their product vanished
identically: $\varphi_l^*(\xx,t=0) \varphi_k(\xx,t=0) \approx 0$. Therefore, column averaging at the initial
instant---which is nothing else but integrating this product over one coordinate---yields zero.
This is also true for any measurement time $t$ because such an averaged expression is nothing else but a scalar
product and as such cannot be changed during a unitary evolution. (Note that this reasoning requires that the
projection of the whole evolution operator onto one dimensional subspace determined by the direction of
propagation of the illuminating light needs still to be unitary, which is true for free evolution but not in general.)

\section{Mott insulator}
\label{sec:mott}

For comparison let us also investigate an ideal Mott insulator, i.e.\ a state with a the same number of
atoms in each well:
\begin{equation}
\left| \psi_{\mbox{\footnotesize mott}} \right\rangle = \left| \rho, \rho, \ldots, \rho \right\rangle
\end{equation}
where $\rho$ is now an integer (and naturally it is still the density of an atomic sample). 
The achievement of the Mott phase is probably the most important experimental result with cold atoms in optical
lattices \cite{Immanuel:Mott}.  Correlation functions in such a
system are straightforward to calculate:
\begin{subequations}
\begin{align}
G^{(1)}_{\mbox{\footnotesize mott}} (\xx) &= \rho \sum_{k} \left| \varphi_k(\xx) \right|^2
\\
G^{(2)}_{\mbox{\footnotesize mott}} (\xx,\xx') &=
\rho (\rho-1) \sum_k \left| \varphi_k(\xx) \right|^2 \left| \varphi_k(\xx') \right|^2 +
\nonumber \\ & +
\rho^2 \sum_{k \neq l} \left| \varphi_k(\xx) \right|^2 \left| \varphi_l(\xx') \right|^2 +
\nonumber \\ & +
\rho^2 \sum_{l} \sum_{k \neq l}
\varphi_{l}^*(\xx) \varphi_k(\xx) \varphi_{l}(\xx') \varphi_{k}^* (\xx')
\label{eq:mott2}
\end{align}
\end{subequations}
These formulas bear a close resemblance to Eq.\ (\ref{eq:random}) and therefore the predictions are quite similar as
well. In
particular, even in a Mott insulating phase one is going to observe interference pattern in a single measurement, 
because its second-order correlation function exhibits the striped structure. The only difference between the Mott
insulator and a state with mixture of Fock states with random phases (or with coherence but in a collapsed phase of
the evolution) lies in the first term in Eq.\ (\ref{eq:mott2}). Consequently, to distinguish between these
states one should perform a measurement with very short measurement times, so that the overlap of wavefunctions of
adjacent sites is vanishing, and look at the diagonal part of the second-order correlation function only, which will
be $G^{(2)}(\xx,\xx) = \omega_0 \rho^2 \sum_k |\varphi_k(\xx)|^4$ for states with mixtures of Fock states at each site
and $G^{(2)}_{\mbox{\footnotesize mott}}(\xx,\xx) = \rho (\rho-1) \sum_k |\varphi_k(\xx)|^4$ 
for Mott insulator. Note that the difference is the
most significant for systems with low densities, in particular $G^{(2)}_{\mbox{\footnotesize mott}}(\xx,\xx)$ vanishes
identically if $\rho=1$.

\section{Conclusions}
\label{sec:conclusions}

We have investigated the recent experiment of collapses and revivals of the matter wave field in terms of
correlation functions. We show that exactly at revival times an interacting system mimics the behaviour of an ideal
gas and all correlation functions factorize. In the collapsed phase of the evolution, on the other hand, the
system effectively behaves as if there was no coherence between the sites of the optical potential. This does not
imply, however, that a smooth density is obtained in a single realization of the experiment: on the
contrary, we show that the second-order correlation function exhibits structure and therefore interference pattern
should be seen in a single measurement. Counter-intuitively, the same conclusions hold also for the Mott insulating
phase.

The fact that the collapse of the matter wave field was seen experimentally at all is due to the column averaging.
We stress the difference between a result of a single measurement and a column averaged single measurement (the
		latter corresponds in fact to a measurement averaged over an ensemble of 2-dimensional experiments). We also
show how to differentiate between the different states via a single measurement.

We investigate the role of the total number of atoms conservation. In particular we identify rare situations in which
predictions of multinomial and coherent states differ. Unsurprisingly, they all correspond to situations with
small number of lattice sites.

\begin{acknowledgments}
We would like to thank I. Bloch for his useful comments.
The work was partially supported by the Subsidy of the Foundation for Polish Science and by the Polish Ministry of
Scientific Research and Information Technology under Grant No.\ PBZ-MIN-008/P03/2003.
\end{acknowledgments}

\end{document}